\documentclass[12pt]{article}
\usepackage{amsmath}
\usepackage{amssymb}

\begin{document}

\bigskip \ 

\bigskip \ 

\begin{center}
\textbf{NEW REFLECTIONS ON HIGHER DIMENSIONAL LINEARIZED GRAVITY}

\smallskip \ 

\smallskip \ 

\smallskip \ 

C. Garc\'{\i}a-Quintero\footnote{%
gqcristhian@uas.edu.mx, cxg163430@utdallas.edu}, A. Ortiz and J. A. Nieto%
\footnote{%
niet@uas.edu.mx, janieto1@asu.edu}

\smallskip \ 

\textit{Facultad de Ciencias F\'{\i}sico-Matem\'{a}ticas de la Universidad
Aut\'{o}noma}

\textit{de Sinaloa, 80010, Culiac\'{a}n, Sinaloa, M\'{e}xico}

\bigskip \ 

\bigskip \ 

\textbf{Abstract}

\smallskip \ 
\end{center}

We make a number of remarks on linearized gravity with cosmological constant
in any dimension, which, we argue, can be useful in a quantum gravity
framework. For this purpose we assume that the background space-time metric
corresponds to the de Sitter or anti-de Sitter space. Moreover, \textit{via}
the graviton mass and the cosmological constant correspondence, we make some
interesting observations, putting special attention on the possible scenario
of \ a graviton-tachyon connection. We compare our proposed formalism with
the Novello and Neves approach.

\bigskip \ 

\bigskip \ 

\bigskip \ 

\bigskip \ 

\bigskip \ 


Pacs numbers: 04.20.Jb, 04.50.-h, 04.60.-m, 11.15.-q

March, 2019

\newpage

\noindent \textbf{1. Introduction}

\smallskip \ 

It is known that there are a number of works relating tachyons with $M$%
-theory [1] (see also Ref. [2] and references therein), including the brane
and anti-brane systems [3], closed-string tachyon condensation [4],
tachyonic instability and tachyon condensation in the $E(8)$ heterotic
string [5], among many others. Part of the motivation of these developments
emerges because it was discovered that the ground state of the bosonic
string is tachyonic [6] and that the spectrum in $AdS/CFT$ [7] can contain a
tachyonic structure.

On the other hand, it is also known that the $(5+5)$-signature and the $%
(1+9) $-signature are common to both type IIA strings and type IIB strings.
In fact, versions of $M$-theory lead to type IIA string theories in
space-time of signatures $(0+10)$, $(1+9)$, $(2+8)$, $(4+6)$ and $(5+5)$,
and to type IIB string theories of signatures $(1+9)$, $(3+7)$ and $(5+5)$
[8]. It is worth mentioning that some of these theories are linked by
duality transformations. So, one wonders whether tachyons may also be
related to the various signatures. In particular, here we are interested to
see the possible relation of tachyons with a space-time of $(4+6)$%
-dimensions. Part of the motivation in the $(4+6)$-signature arises from the
observation that $(4+6)=(1+4)+(3+2)$. This means that the world of $(4+6)$%
-dimensions can be splitted into a de Sitter world of $(1+4)$-dimensions and
an anti-de Sitter world of $(3+2)$-dimensions. Moreover, looking the $(4+6)$%
-world from the perspective of $(3+7)$-dimensions obtained by
compactifying-uncompactifying prescription such that $4\rightarrow 3$ and $%
6\rightarrow 7$, one can associate with the $3$ and $7$ dimensions of the $%
(3+7)$-world with a $S^{3}$ and $S^{7}$, respectively, which are two of the
parallelizable spheres; the other it is $S^{1}$. As it is known these
spheres are related to both Hopf maps and division algebras (see Ref. [9]
and references therein).

In this work, we develop a formalism that allows us to address the $(4+6)$%
-dimensional world \textit{via} linearized gravity. In this case, one starts
assuming the Einstein field equations with cosmological constant $\Lambda $
in $(4+6)$-dimensions and develops the formalism considering a linearized
metric in such equations. We note that the result is deeply related to the
cosmological constant $\Lambda \lessgtr 0$ sign. In fact, one should
remember that in $(1+4)$-dimensions, $\Lambda $ is positive, while in $(3+2)$%
-dimensions, $\Lambda $ is negative. At the level of linearized gravity, one
searches for the possibility of associating these two different signs of $%
\Lambda $ with tachyons. This leads us to propose a unified tachyonic
framework in $(4+6)$-dimensions which includes these two separate cases of $%
\Lambda $. Moreover, we argue that our formalism may admit a possible
connection with the increasing interesting proposal of duality in linearized
gravity (see Refs. [10-12] and references therein).

In order to achieve our goal, we first introduce, in a simple context, the
tachyon theory. Secondly, in a novel form we develop the de Sitter and
anti-de Sitter space-times formalism, clarifying the meaning of the main
constraints. Moreover, much work allows us to describe a new formalism for
higher dimensional linearized gravity. Our approach is focused on the
space-time signature in any dimension and in particular in $(4+6)$%
-dimensions.

A further motivation of our approach may emerge from the recent direct
detections of gravitational waves\ [13-15]. According to this detection the
upper bound of the graviton mass is $m_{g}\leq 1.3\times 10^{-58}kg$ [15].
Since in our computations the mass and the cosmological constant are
proportional, such an upper bound must also be reflected in the cosmological
constant value.

Technically, this work is structured as follows. In section 2, we make a
simple introduction of tachyon theory. In section 3, we discuss a possible
formalism for the de Sitter and anti-de Sitter space-times. In section 4, we
develop the most general formalism of higher dimensional linearized gravity
with cosmological constant. In section 5, we establish a novel approach for
considering the constraints that determine the de Sitter and anti-de Sitter
space. In section 6, we associate the concept of tachyons with higher
dimensional linearized gravity. In section 7, we develop linearized gravity
with cosmological constant in $(4+6)$-dimensions. We add an appedix A in
attempt to further clarify the negative mass squared term-tachyon
association. Finally, in section 8, we make some final remarks.

\bigskip \ 

\noindent \textbf{2. Special relativity and the signature of the space-time}

\smallskip \ 

Let us start considering the well known time dilatation formula

\begin{equation}
dt=\frac{d\tau }{\sqrt{1-\frac{v^{2}}{c^{2}}}}.  \label{1}
\end{equation}%
Here, $\tau $ is the proper time, $v^{2}\equiv $ $(\frac{dx}{dt})^{2}+(\frac{%
dy}{dt})^{2}+(\frac{dz}{dt})^{2}$ is the velocity of the object and $c$
denotes the speed of light. Of course, the expression (1) makes sense over
the real numbers only if one assumes $v<c$. It is straightforward to see
that (1) leads to the line element

\begin{equation}
ds^{2}=-c^{2}d\tau ^{2}=-c^{2}dt^{2}+dx^{2}+dy^{2}+dz^{2}.  \label{2}
\end{equation}%
In tensorial notation one may write (2) as

\begin{equation}
ds^{2}=-c^{2}d\tau ^{2}=\eta _{\mu \nu }^{(+)}dx^{\mu }dx^{\nu },  \label{3}
\end{equation}%
where the indices $\mu ,\nu $ take values in the set $\{1,2,3,4\}$, $%
x^{1}=ct $, $x^{2}=x$, $x^{3}=y$ and $x^{4}=z$. Moreover, $\eta _{\mu \nu
}^{(+)}$ denotes the flat Minkowski metric with associated signature $%
(-1,+1,+1,+1)$. Usually, one says that such a signature represents a world
of $(1+3)$-dimensions.

If one now defines the linear momentum

\begin{equation}
p^{\mu }=m_{0}^{(+)}\frac{dx^{\mu }}{d\tau },  \label{4}
\end{equation}%
with $m_{0}^{(+)}\neq 0$ a constant, one sees that (3) implies

\begin{equation}
p^{\mu }p^{\nu }\eta _{\mu \nu }^{(+)}+m_{0}^{(+)2}c^{2}=0.  \label{5}
\end{equation}%
Of course, $m_{0}^{(+)}$ plays the role of the rest mass of the object. This
is because setting $p^{i}=0$, with $i\in \{2,3,4\}$, in the rest frame and
defining $E=cp^{1}$, the constraint (5) leads to the famous formula $E=\pm
m_{0}c^{2}$.

Let us follow similar steps, but instead of starting with the expression
(1), one now assumes the formula

\begin{equation}
d\lambda =\frac{d\xi }{\sqrt{\frac{u^{2}}{c^{2}}-1}},  \label{6}
\end{equation}%
where $u^{2}=(\frac{dw}{d\xi })^{2}+(\frac{d\rho }{d\xi })^{2}+(\frac{d\zeta 
}{d\xi })^{2}$. Note that in this case one has the condition $u>c$. Here, in
order to emphasize the differences between (1) and (6), we are using a
different notation. Indeed, the notation used in (1) and (6) is introduced
in order to establish an eventual connection with $(4+6)$-dimensions. From
(6) one obtains

\begin{equation}
d\mathit{s}^{2}=-c^{2}d\xi ^{2}=+c^{2}d\lambda ^{2}-dw^{2}-d\rho ^{2}-d\zeta
^{2}.  \label{7}
\end{equation}%
In tensorial notation, one may write (7) as

\begin{equation}
ds^{2}=-c^{2}d\xi ^{2}=\eta _{\mu \nu }^{(-)}dy^{\mu }dy^{\nu },  \label{8}
\end{equation}%
where $y^{1}=c\lambda $, $y^{2}=w$, $y^{3}=\rho $ and $y^{4}=\zeta $.
Moreover, $\eta _{\mu \nu }^{(-)}$ denotes the flat Minkowski metric with
associated signature $(+1,-1,-1,-1)$. One says that this signature
represents a world of $(3+1)$-dimensions.

If one now defines the linear momentum

\begin{equation}
\mathcal{P}^{\mu }=m_{0}^{(-)}\frac{dy^{\mu }}{d\xi },  \label{9}
\end{equation}%
with $m_{0}^{(-)}\neq 0$ a constant, one sees that (9) implies

\begin{equation}
\mathcal{P}^{\mu }\mathcal{P}^{\nu }\eta _{\mu \nu
}^{(-)}+m_{0}^{(-)2}c^{2}=0.  \label{10}
\end{equation}%
Since, $u>c$ one observes that in this case the constraint (10) corresponds
to a tachyon system with mass $m_{0}^{(-)}$.

Now, for the case of ordinary matter, if one wants to quantize, one starts
promoting $p^{\mu }$ as an operator identifying $\hat{p}^{\mu }=-i\partial
^{\mu }$. Thus, at the quantum level (5) becomes%
\begin{equation}
(-\partial ^{\mu }\partial ^{\nu }\eta _{\mu \nu
}^{(+)}+m_{0}^{(+)2})\varphi =0.  \label{11}
\end{equation}%
It is important to mention that here we are using a coordinate
representation for $\varphi $ in the sense that $\varphi (x^{\mu })=<x^{\mu
}|\varphi >$.

By defining the d'Alembert operator ${\square ^{(+)}}^{2}=\eta _{\mu \nu
}^{(+)}\partial ^{\mu }\partial ^{\nu }$ one notes that last equation reads

\begin{equation}
{(\square ^{(+)}}^{2}-m_{0}^{(+)2})\varphi =0.  \label{12}
\end{equation}%
Analogously, in the constraint (10) one promotes the momentum $\mathcal{P}%
^{\mu }$ as an operator $\mathcal{\hat{P}}^{\mu }=-i\partial ^{\mu }$ and
using $\eta _{\mu \nu }^{(+)}=-\eta _{\mu \nu }^{(-)}$, the expression (10)
yields

\begin{equation}
({\square ^{(+)}}^{2}+m_{0}^{(-)2})\varphi =0.  \label{13}
\end{equation}%
The last two expressions are Klein-Gordon type equations for ordinary matter
and tachyonic systems, respectively. In fact, these two equations will play
an important role in the analysis in section 6, concerning linearized
gravity with positive and negative cosmological constant.

\bigskip \ 

\noindent \textbf{3. Clarifying de Sitter and anti-de Sitter space-time}

\smallskip \ 

Let us start with the constraint%
\begin{equation}
x^{i}x^{j}\eta _{ij}^{(+)}+(x^{d+1})^{2}=r_{0}^{2},  \label{14}
\end{equation}%
where $\eta _{ij}^{(+)}=diag(-1,1,...,1)$ is the Minkowski metric and the $i$
index goes from $1$ to $d$. and $r_{0}^{2}$ is a positive constant. The line
element is given by

\begin{equation}
ds^{2}\equiv dx^{A}dx^{B}\eta _{AB}=dx^{i}dx^{j}\eta
_{ij}^{(+)}+(dx^{d+1})^{2}.  \label{15}
\end{equation}%
It is not difficult to see that the corresponding Christoffel symbols and
the Riemann tensor are given by

\begin{equation}
\Gamma _{kl}^{i}=\frac{g_{kl}x^{i}}{r_{0}^{2}}  \label{16}
\end{equation}%
and

\begin{equation}
R_{ijkl}=\frac{1}{r_{0}^{2}}(g_{ik}g_{jl}-g_{il}g_{jk}),  \label{17}
\end{equation}%
respectively.

Here, the metric $g_{ij}$ is given by

\begin{equation}
g_{ij}=\eta _{ij}^{(+)}+\frac{x_{i}x_{j}}{(r_{0}^{2}-x^{r}x^{s}\eta
_{rs}^{(+)})}.  \label{18}
\end{equation}%
It is worth mentioning that one can even consider a flat metric $\eta
_{ij}=diag(-1,-1,....,1,1)$, with $t$-times and $s$-space coordinates and
analogue developments leads to the formulas (14)-(18).\newline

Of course, the line element associated with the metric (18) is

\begin{equation}
ds^{2}\equiv (\eta _{ij}^{(+)}+\frac{x_{i}x_{j}}{(r_{0}^{2}-x^{r}x^{s}\eta
_{rs}^{(+)})})dx^{i}dx^{j},  \label{19}
\end{equation}%
which in spherical coordinates becomes%
\begin{equation}
ds^{2}\equiv -(1-\frac{r^{2}}{r_{0}^{2}})dt^{2}+\frac{dr^{2}}{(1-\frac{r^{2}%
}{r_{0}^{2}})}+r^{2}d\Omega ^{d-2}.  \label{20}
\end{equation}%
Here, one is assuming that $x^{m}x^{n}\eta _{mn}^{(+)}=-x^{1}x^{1}+r^{2}$,
where $r^{2}=x^{a}x^{b}\delta _{ab}$, with $a,b$ running from $2$ to $d$.
Moreover, $d\Omega ^{d-2}$ is a volume element in $d-2$ dimensions. The
expression (20) is, of course, very useful when one considers black-holes or
cosmological models in the de Sitter space (or anti-de Sitter space).

In the anti-de\ Sitter case, instead of starting with the formula (14) one
considers the constraint is $x^{i}x^{j}\eta
_{ij}^{(+)}-(x^{d+1})^{2}=-r_{0}^{2}$. This constraint will play an
important role in section 5.

\bigskip \newpage

\noindent \textbf{4. Linearized gravity with cosmological constant in any
dimension}

\smallskip \ 

Although in the literature there are similar computations [16], the
discussion of this section seems to be new, in sense that it is extended to
any background metric in higher dimensions. Usually, one starts linearized
gravity by writing the metric of the space-time $g_{\mu \nu }=g_{\mu \nu
}(x^{\alpha })$ as

\begin{equation}
g_{\mu \nu }=\eta _{\mu \nu }+h_{\mu \nu },  \label{21}
\end{equation}%
where $\eta _{\mu \nu }=diag(-1,-1,....1,1)$ is the Minkowski metric, with $%
t $-times and $s$-space coordinates, and $h_{\mu \nu }$ is a small
perturbation. Therefore, the general idea is to keep only with the first
order terms in $h_{\mu \nu }$, in the Einstein field equations.

Here, we shall replace the Minkowski metric $\eta _{\mu \nu }$ by a general
background metric denoted by $g_{\mu \nu }^{(0)}$. At the end we shall
associate $g_{\mu \nu }^{(0)}$ with the de Sitter or anti-de Sitter space.
So, the analogue of (21) becomes

\begin{equation}
g_{\mu \nu }=g_{\; \; \; \; \mu \nu }^{(0)}+h_{\mu \nu }.  \label{22}
\end{equation}%
The inverse of $g_{\mu \nu }$ is

\begin{equation}
g^{\mu \nu }=g^{(0){\mu \nu }}-h^{\mu \nu }.  \label{23}
\end{equation}

\noindent Here, (23) is the inverse metric of (22) at first order in $h_{\mu
\nu}$. Also, the metric $g_{\; \; \; \; \mu \nu }^{(0)}$ is used to raise
and lower indices. Therefore, neglecting the terms of second order in $%
h_{\mu \nu }$ one finds that the Christoffel symbols can be written as

\begin{equation}
\Gamma _{\mu \nu }^{\lambda }=\Gamma _{\; \; \; \; \mu \nu }^{(0)\lambda
}+\Sigma _{\mu \nu ,}^{\lambda }  \label{24}
\end{equation}%
where $\Gamma _{\; \; \; \; \mu \nu }^{(0)\lambda }$ are the Christoffel
symbols associated with $g_{\; \; \; \; \mu \nu }^{(0)}$ and $\Sigma _{\mu
\nu }^{\lambda }$ is given by

\begin{equation}
\Sigma _{\mu \nu }^{\lambda }\equiv \frac{1}{2}g^{(0)\lambda \alpha }(%
\mathcal{D}_{\nu }h_{\alpha \mu }+\mathcal{D}_{\mu }h_{\nu \alpha }-\mathcal{%
D}_{\alpha }h_{\mu \nu }).  \label{25}
\end{equation}%
Here, the symbol $\mathcal{D}_{\mu }$ denotes covariant derivative with
respect the metric $g_{\; \; \; \; \mu \nu }^{(0)}$.

Similarly, one obtains that at first order in $h_{\mu \nu }$, the Riemann
tensor becomes

\begin{equation}
R_{\; \; \nu \alpha \beta }^{\mu }=R_{\; \; \; \; \; \; \; \nu \alpha \beta
}^{(0)\mu }+\mathcal{D}_{\alpha }\Sigma _{\; \; \nu \beta }^{\mu }-\mathcal{D%
}_{\beta }\Sigma _{\; \; \nu \alpha }^{\mu },  \label{26}
\end{equation}%
which can be rewritten as

\begin{equation}
R_{\mu \nu \alpha \beta }=R_{\mu \nu \alpha \beta }^{(0)}+\mathcal{D}%
_{\alpha }\Sigma _{\nu \beta \mu }-\mathcal{D}_{\beta }\Sigma _{\nu \alpha
\mu }+R_{\; \; \; \; \; \; \; \nu \alpha \beta }^{(0)\sigma }h_{\sigma \mu },
\label{27}
\end{equation}%
where

\begin{equation}
\Sigma _{\mu \nu \alpha }\equiv \frac{1}{2}(\mathcal{D}_{\nu }h_{\alpha \mu
}+\mathcal{D}_{\mu }h_{\alpha \nu }-\mathcal{D}_{\alpha }h_{\mu \nu }).
\label{28}
\end{equation}%
Then, using the definition (28), the Riemann tensor becomes

\begin{equation}
\begin{array}{c}
R_{\mu \nu \alpha \beta }=\frac{1}{2}(\mathcal{D}_{\alpha }\mathcal{D}%
_{\beta }h_{\mu \nu }-\mathcal{D}_{\beta }\mathcal{D}_{\alpha }h_{\mu \nu }+%
\mathcal{D}_{\alpha }\mathcal{D}_{\nu }h_{\beta \mu }-\mathcal{D}_{\beta }%
\mathcal{D}_{\nu }h_{\alpha \mu } \\ 
\\ 
+\mathcal{D}_{\beta }\mathcal{D}_{\mu }h_{\nu \alpha }-\mathcal{D}_{\alpha }%
\mathcal{D}_{\mu }h_{\nu \beta })+R_{\mu \nu \alpha \beta }^{(0)}+R_{\; \;
\; \; \; \; \; \nu \alpha \beta }^{(0)\sigma }h_{\sigma \mu }.%
\end{array}
\label{29}
\end{equation}%
Note that in this case the covariant derivatives $\mathcal{D}_{\mu }$ do not
commute as is the case of the ordinate partial derivatives $\partial _{\mu }$
in a Minkowski space-time background.

One can show that the term $\mathcal{D}_{\alpha }\mathcal{D}_{\beta }h_{\mu
\nu }-\mathcal{D}_{\beta }\mathcal{D}_{\alpha }h_{\mu \nu }$ leads to

\begin{equation}
\mathcal{D}_{\alpha }\mathcal{D}_{\beta }h_{\mu \nu }-\mathcal{D}_{\beta }%
\mathcal{D}_{\alpha }h_{\mu \nu }=-h_{\lambda \mu }R_{\; \; \; \; \; \; \;
\nu \alpha \beta }^{(0)\lambda }-h_{\lambda \nu }R_{\; \; \; \; \; \; \; \mu
\alpha \beta }^{(0)\lambda }.  \label{30}
\end{equation}%
Then using (29), (30) and properties of the Riemann tensor, one can rewrite $%
R_{\mu \nu \alpha \beta }$ as

\begin{equation}
\begin{array}{c}
R_{\mu \nu \alpha \beta }=R_{\mu \nu \alpha \beta }^{(0)}+\frac{1}{2}%
(h_{\lambda \alpha }R_{\; \; \; \; \; \; \; \mu \beta \nu }^{(0)\lambda
}-h_{\lambda \beta }R_{\; \; \; \; \; \; \; \mu \alpha \nu }^{(0)\lambda
}-h_{\lambda \nu }R_{\; \; \; \; \; \; \; \mu \alpha \beta }^{(0)\lambda }
\\ 
\\ 
+\mathcal{D}_{\nu }\mathcal{D}_{\alpha }h_{\mu \beta }-\mathcal{D}_{\nu }%
\mathcal{D}_{\beta }h_{\mu \alpha }+\mathcal{D}_{\beta }\mathcal{D}_{\mu
}h_{\nu \alpha }-\mathcal{D}_{\alpha }\mathcal{D}_{\mu }h_{\nu \beta }).%
\end{array}
\label{31}
\end{equation}%
Multiplying (31) by $g^{\mu \nu }$, as given in (23), leads to the Ricci
tensor

\begin{equation}
\begin{array}{c}
R_{\mu \nu }=R_{\; \; \; \; \mu \nu }^{(0)}+\frac{1}{2}(h_{\lambda \nu
}R_{\; \; \; \; \; \; \; \mu }^{(0)\lambda }+h_{\lambda \mu }R_{\; \; \; \;
\; \; \; \nu }^{(0)\lambda })-h^{\alpha \beta }R_{\alpha \mu \beta \nu
}^{(0)} \\ 
\\ 
+\frac{1}{2}\left( \mathcal{D}_{\mu }\mathcal{D}^{\alpha }h_{\alpha \nu }+%
\mathcal{D}_{\nu }\mathcal{D}^{\alpha }h_{\alpha \mu }-\mathcal{D}_{\mu }%
\mathcal{D}_{\nu }h-\mathcal{D}^{\alpha }\mathcal{D}_{\alpha }h_{\mu \nu
}\right) .%
\end{array}
\label{32}
\end{equation}%
Thus, the scalar curvature $R=g^{\mu \nu }R_{\mu \nu }$ becomes

\begin{equation}
R=R^{(0)}+\mathcal{D}^{\alpha }\mathcal{D}^{\beta }h_{\alpha \beta }-%
\mathcal{D}^{\alpha }\mathcal{D}_{\alpha }h-h^{\alpha \beta }R_{\; \; \; \;
\alpha \beta }^{(0)}.  \label{33}
\end{equation}%
Now one can use (32) and (33) in the Einstein gravitational field equations
with a cosmological constant

\begin{equation}
R_{\mu \nu }-\frac{1}{2}g_{\mu \nu }R+\Lambda g_{\mu \nu }=0.  \label{34}
\end{equation}%
When one sets $g_{\mu \nu }^{(0)}$ as a de Sitter (or anti-de Sitter)
background one obtains

\begin{equation}
\begin{array}{c}
\mathcal{D}_{\mu }\mathcal{D}_{\nu }h+\mathcal{D}^{\alpha }\mathcal{D}%
_{\alpha }h_{\mu \nu }-\mathcal{D}_{\mu }\mathcal{D}^{\alpha }h_{\alpha \nu
}-\mathcal{D}_{\nu }\mathcal{D}^{\alpha }h_{\alpha \mu } \\ 
\\ 
+g_{\; \; \; \; \mu \nu }^{(0)}(\mathcal{D}^{\alpha }\mathcal{D}^{\beta
}h_{\alpha \beta }-\mathcal{D}^{\alpha }\mathcal{D}_{\alpha }h) \\ 
\\ 
-\frac{2}{r_{0}^{2}}h_{\mu \nu }-\frac{(d-3)}{2r_{0}^{2}}hg_{\; \; \; \; \mu
\nu }^{(0)}=0.%
\end{array}
\label{35}
\end{equation}%
As it is commonly done, in linearized gravity in four dimensions, one shall
define $\bar{h}_{\mu \nu }=h_{\mu \nu }-\frac{1}{2}g_{\; \; \; \; \mu \nu
}^{(0)} $. Therefore, substituting this expression for $\bar{h}_{\mu \nu }$
in (35), fixing the Lorentz gauge $\mathcal{D}^{\nu }\bar{h}_{\mu \nu }=0$
and assuming the trace $\bar{h}=0$, one finally gets

\begin{equation}
\square ^{2}\bar{h}_{\mu \nu }-\frac{4\Lambda }{(d-2)(d-1)}\bar{h}_{\mu \nu
}=0,  \label{36}
\end{equation}

\noindent where $d$ is the dimension of the space-time. It is important to
observe that in (36), $\square ^{2}=\eta _{\mu \nu }^{(+)}\partial ^{\mu
}\partial ^{\nu }$ is now generalized to the form $\square ^{2}=g_{\; \; \;
\; \mu \nu }^{(0)}\mathcal{D}^{\mu }\mathcal{D}^{\nu }$.

At this point, considering the $(4+6)$-signature (which can be splitted into
a de Sitter and an anti-de Sitter space according to $(4+6)=(1+4)+(3+2)$)
one has to set $d=8$ since there are two constraints, one given by the de
Sitter world and another from the anti-de\ Sitter world. Consequently, the
equation (36) becomes

\begin{equation}
\square ^{2}\bar{h}_{\mu \nu }-\frac{2}{21}\Lambda \bar{h}_{\mu \nu }=0.
\label{37}
\end{equation}%
One recognizes this expression as the equation of a gravitational wave in $%
d=8$.

\bigskip \ 

\noindent \textbf{5. Constraints in de Sitter and anti de Sitter space}

\smallskip \ 

When one considers the de Sitter space, one assumes the constraint (14).
However, one may notice that actually there are eight possible constraints
corresponding to the two metrics $\eta _{\mu \nu }^{(+)}$ and $\eta _{\mu
\nu }^{(-)}$ mentioned in section 3. In fact, for the metric $\eta _{\mu \nu
}^{(+)}$ one has following possibilities:

\begin{equation}
x^{\mu }x^{\nu }\eta _{\mu \nu }^{(+)}+(x^{d+1})^{2}=r_{0}^{2},  \label{38}
\end{equation}

\begin{equation}
x^{\mu }x^{\nu }\eta _{\mu \nu }^{(+)}-(x^{d+1})^{2}=r_{0}^{2},  \label{39}
\end{equation}

\begin{equation}
x^{\mu }x^{\nu }\eta _{\mu \nu }^{(+)}+(x^{d+1})^{2}=-r_{0}^{2}  \label{40}
\end{equation}%
and

\begin{equation}
x^{\mu }x^{\nu }\eta _{\mu \nu }^{(+)}-(x^{d+1})^{2}=-r_{0}^{2}.  \label{41}
\end{equation}%
While for the metric $\eta _{\mu \nu }^{(-)}$ one finds 
\begin{equation}
x^{\mu }x^{\nu }\eta _{\mu \nu }^{(-)}+(x^{d+1})^{2}=r_{0}^{2},  \label{42}
\end{equation}

\begin{equation}
x^{\mu }x^{\nu }\eta _{\mu \nu }^{(-)}-(x^{d+1})^{2}=r_{0}^{2},  \label{43}
\end{equation}

\begin{equation}
x^{\mu }x^{\nu }\eta _{\mu \nu }^{(-)}+(x^{d+1})^{2}=-r_{0}^{2}  \label{44}
\end{equation}%
and

\begin{equation}
x^{\mu }x^{\nu }\eta _{\mu \nu }^{(-)}-(x^{d+1})^{2}=-r_{0}^{2}.  \label{45}
\end{equation}%
Now, since one has the relation $\  \eta _{\mu \nu }^{(+)}=-\eta _{\mu \nu
}^{(-)}$, one sees that two sets of constraints (38)-(41) and (42)-(45) are
equivalents. Hence, we shall focus only on the first set of constraints
(38)-(41). Let us now rewrite (39) and (40) as

\begin{equation}
x^{\mu }x^{\nu }\eta _{\mu \nu }^{(+)}=[(x^{d+1})^{2}+r_{0}^{2}]  \label{46}
\end{equation}%
and

\begin{equation}
x^{\mu }x^{\nu }\eta _{\mu \nu }^{(+)}=-[(x^{d+1})^{2}+r_{0}^{2}].
\label{47}
\end{equation}%
Observe that one may assume that $x^{\mu }x^{\nu }\eta _{\mu \nu
}^{(+)}\lessgtr 0$, with $\eta _{\mu \nu }^{(+)}=(-1,1,...,1)$. So, since
right hand side of (46) is strictly positive, by consistency this constraint
must be omitted. Similarly since the right hand side of (47) is strictly
negative then this constraint must also be omitted. Thus, one only needs to
consider the two constraints, (38) and (41) which corresponds to polynomial
equations over the reals whose set of solutions can be associated with
classically algebraic varieties. When this constraints are used at the level
of line element, one discovers that they can be associated with manifolds
for the de Sitter and anti-de Sitter space-time. It turns out that the
constraints (38) and (41) can be rewritten as

\begin{equation}
x^{A}x^{B}\eta _{AB}^{(+)}=r_{0}^{2}  \label{48}
\end{equation}%
and

\begin{equation}
x^{A}x^{B}\eta _{AB}^{(-)}=\rho _{0}^{2},  \label{49}
\end{equation}%
where

\begin{equation}
\eta _{AB}^{(+)}=(-1,1,...,1,1)  \label{50}
\end{equation}%
and

\begin{equation}
\eta _{AB}^{(-)}=(1,-1,-1,...,-1,1).  \label{51}
\end{equation}%
Here, one is assuming that (49) allows for a different radius $\rho _{0}$.
This is useful for emphasizing that $r_{0}^{2}$ refers to the de Sitter
world and $\rho _{0}^{2}$ to the anti-de Sitter world. Note that using $\rho
_{0}^{2}$ the expression (49) can also be rewritten as

\begin{equation}
x^{i}x^{j}\eta _{ij}^{(-)}+(x^{d+1})^{2}=\rho _{0}^{2}.  \label{52}
\end{equation}%
Now, using (50) and (51) one can write the line elements in the form

\begin{equation}
ds_{(+)}^{2}=dx^{A}dx^{B}\eta _{AB}^{(+)}=dx^{\mu }dx^{\nu }\eta _{\mu \nu
}^{(+)}+(dx^{d+1})^{2}  \label{53}
\end{equation}%
and

\begin{equation}
ds_{(-)}^{2}=dx^{A}dx^{B}\eta _{AB}^{(-)}=dx^{\mu }dx^{\nu }\eta _{\mu \nu
}^{(-)}+(dx^{d+1})^{2}.  \label{54}
\end{equation}

From (38) one obtains

\begin{equation}
x^{d+1}=\pm (r^{2}_{0}-x^{\alpha }x^{\beta }\eta _{\alpha \beta }^{(+)})^{%
\frac{1}{2}}.  \label{55}
\end{equation}

So, differentiating (55) one obtains

\begin{equation}
dx^{d+1}=\mp \frac{x_{\mu }}{r^{2}_{0}-x^{\alpha }x^{\beta }\eta _{\alpha
\beta }^{(+)}}dx^{\mu }.  \label{56}
\end{equation}%
Similarly, from (52) one gets

\begin{equation}
dx^{d+1}=\mp \frac{x_{\mu }}{\rho _{0}^{2}-x^{\alpha }x^{\beta }\eta
_{\alpha \beta }^{(-)}}dx^{\mu }.  \label{57}
\end{equation}%
Hence, with the help of (56) and (57), one can rewrite (53) and (54) as

\begin{equation}
ds_{(+)}^{2}=\left( \eta _{\mu \nu }^{(+)}+\frac{x_{\mu }x_{\nu }}{%
r^{2}_{0}-x^{\alpha }x^{\beta }\eta _{\alpha \beta }^{(+)}}\right) dx^{\mu
}dx^{\nu }  \label{58}
\end{equation}%
and

\begin{equation}
ds_{(-)}^{2}=\left( \eta _{\mu \nu }^{(-)}+\frac{x_{\mu }x_{\nu }}{\rho
_{0}^{2}-x^{\alpha }x^{\beta }\eta _{\alpha \beta }^{(-)}}\right) dx^{\mu
}dx^{\nu },  \label{59}
\end{equation}%
respectively.

Thus, one learns that the metrics associated with (58) and (59) are

\begin{equation}
g_{\mu \nu }^{(+)}=\eta _{\mu \nu }^{(+)}+\frac{x_{\mu }x_{\nu }}{%
r^{2}_{0}-x^{\alpha }x^{\beta }\eta _{\alpha \beta }^{(+)}}  \label{60}
\end{equation}%
and

\begin{equation}
g_{\mu \nu }^{(-)}=\eta _{\mu \nu }^{(-)}+\frac{x_{\mu }x_{\nu }}{\rho
_{0}^{2}-x^{\alpha }x^{\beta }\eta _{\alpha \beta }^{(-)}},  \label{61}
\end{equation}%
respectively.

Using (60) and (61) one sees that according to (17) the Riemann tensors $%
R_{\mu \nu \alpha \beta }^{(+)}$ and $R_{\mu \nu \alpha \beta }^{(-)}$ become

\begin{equation}
R_{\mu \nu \alpha \beta }^{(+)}=\frac{1}{r_{0}^{2}}\left( g_{\mu \alpha
}^{(+)}g_{\nu \beta }^{(+)}-g_{\mu \beta }^{(+)}g_{\nu \alpha }^{(+)}\right)
\label{62}
\end{equation}%
and

\begin{equation}
R_{\mu \nu \alpha \beta }^{(-)}=-\frac{1}{\rho _{0}^{2}}\left( g_{\mu \alpha
}^{(-)}g_{\nu \beta }^{(-)}-g_{\mu \beta }^{(-)}g_{\nu \alpha }^{(-)}\right)
.  \label{63}
\end{equation}%
The corresponding curvature scalars associated with (62) and (63) are

\begin{equation}
R^{(+)}=\frac{d(d-1)}{r_{0}^{2}}  \label{64}
\end{equation}%
and

\begin{equation}
R^{(-)}=-\frac{d(d-1)}{\rho _{0}^{2}}.  \label{65}
\end{equation}

Now, let us consider the Einstein gravitational field equation (see eq. (34))

\begin{equation}
R_{\mu \nu }^{(+)}-\frac{1}{2}g_{\mu \nu }^{(+)}R^{(+)}+\Lambda ^{(+)}g_{\mu
\nu }^{(+)}=0,  \label{66}
\end{equation}%
for $g_{\mu \nu }^{(+)}$. Multiplying this expression by $g^{\mu \nu (+)}$
one sees that (66) leads to

\begin{equation}
R^{(+)}-\frac{1}{2}dR^{(+)}+\Lambda ^{(+)}d=0.  \label{67}
\end{equation}%
Solving (67) for $\Lambda ^{(+)}$ leads to

\begin{equation}
\Lambda ^{(+)}=\frac{(d-2)(d-1)}{2r_{0}^{2}},  \label{68}
\end{equation}%
where the equation (64) was used.

In analogous way, by considering the Einstein gravitational field equations
for $g_{\mu \nu }^{(-)}$

\begin{equation}
R_{\mu \nu }^{(-)}-\frac{1}{2}g_{\mu \nu }^{(-)}R^{(-)}+\Lambda ^{(-)}g_{\mu
\nu }^{(-)}=0,  \label{69}
\end{equation}%
one obtains

\begin{equation}
\Lambda ^{(-)}=-\frac{(d-2)(d-1)}{2\rho _{0}^{2}}.  \label{70}
\end{equation}%
Note that, since $\Lambda ^{(-)}$ refers to the anti-de Sitter space, (70)
agrees with the fact that $\Lambda ^{(-)}<0$.

\bigskip \ 

\noindent \textbf{6. The signature of the space-time in linearized gravity}

\smallskip \ 

In the previous section, the Einstein gravitational field equations were
considered for $g_{\mu \nu }^{(+)}$ and $g_{\mu \nu }^{(-)}$. Such equations
lead us to find a relations for $\Lambda ^{(+)}$ and $\Lambda ^{(-)}$. Now,
if one substitutes the equations (68) and (70) into (36) one obtains

\begin{equation}
\left( ^{(+)}\square ^{2}-\frac{4\Lambda ^{(+)}}{(d-2)(d-1)}\right) \bar{h}%
_{\mu \nu }^{(+)}=0,  \label{71}
\end{equation}%
and

\begin{equation}
\left( ^{(-)}\square ^{2}+\frac{4\Lambda ^{(-)}}{(d-2)(d-1)}\right) \bar{h}%
_{\mu \nu }^{(-)}=0.  \label{72}
\end{equation}%
Here, $^{(\pm )}\square ^{2}=g_{\; \; \; \; \mu \nu }^{(\pm )(0)}\mathcal{D}%
^{\mu }\mathcal{D}^{\nu }$.

Let us now to consider, in the context of linearized gravity, the vielbein
formalism for $g_{\mu \nu }^{(+)}$ and $g_{\mu \nu }^{(-)}$. One introduces
the vielbein field $e_{\mu }^{a}$ and writes

\begin{equation}
g_{\mu \nu }^{(\pm )}=e_{\mu }^{a}e_{\nu }^{b}\eta _{ab}^{(\pm )},
\label{73}
\end{equation}%
where

\begin{equation}
e_{\mu }^{a}=b_{\mu }^{a}+h_{\mu }^{a}.  \label{74}
\end{equation}%
If one replaced (74) into (73), the metric $g_{\mu \nu }^{(\pm )}$ becomes

\begin{equation}
g_{\mu \nu }^{(\pm )}=(b_{\mu }^{a}+h_{\mu }^{a})(b_{\nu }^{b}+h_{\nu
}^{b})\eta _{ab}^{(\pm )}.  \label{75}
\end{equation}%
Thus, one obtains

\begin{equation}
g_{\mu \nu }^{(\pm )}=b_{\mu }^{a}b_{\nu }^{b}\eta _{ab}^{(\pm )}+b_{\mu
}^{a}h_{\nu }^{b}\eta _{ab}^{(\pm )}+h_{\mu }^{a}b_{\nu }^{b}\eta
_{ab}^{(\pm )}+h_{\mu }^{a}h_{\nu }^{b}\eta _{ab}^{(\pm )}.  \label{76}
\end{equation}%
Since one is assuming that $h_{\mu }^{a}\ll 1$ then $h_{\mu }^{a}h_{\nu
}^{b}\eta _{ab}^{(\pm )}\sim 0$ and therefore (76) is reduced to

\begin{equation}
g_{\mu \nu }^{(\pm )}=b_{\mu }^{a}b_{\nu }^{b}\eta _{ab}^{(\pm )}+b_{\mu
}^{a}h_{\nu }^{b}\eta _{ab}^{(\pm )}+h_{\mu }^{a}b_{\nu }^{b}\eta
_{ab}^{(\pm )}.  \label{77}
\end{equation}%
If one establishes the identifications $g_{\; \; \; \; \mu \nu }^{(\pm
)(0)}=b_{\mu }^{a}b_{\nu }^{b}\eta _{ab}^{(\pm )}$ and $h_{\mu \nu }^{(\pm
)}=h_{\mu }^{a}b_{\nu }^{b}\eta _{ab}^{(\pm )}+b_{\mu }^{a}h_{\nu }^{b}\eta
_{ab}^{(\pm )}$ one obtains

\begin{equation}
g_{\mu \nu }^{(\pm )}=g_{\; \; \; \; \mu \nu }^{(\pm )(0)}+h_{\mu \nu
}^{(\pm )},  \label{78}
\end{equation}%
which is the expression (22) but with the signatures $+$ or $-$ in $g_{\; \;
\; \; \mu \nu }^{(\pm )(0)}$ and $h_{\mu \nu }^{(\pm )}$ identified.

Now, we shall compare the equations (71) and (72) with (12) and (13),
respectively. Since $\Lambda ^{(+)}>0$ and $\Lambda ^{(-)}<0$ one can
introduces the two mass terms

\begin{equation}
m^{(+)2}=\frac{4\Lambda ^{(+)}}{(d-2)(d-1)}  \label{79}
\end{equation}%
and

\begin{equation}
m^{(-)2}=-\frac{4\Lambda ^{(-)}}{(d-2)(d-1)}.  \label{80}
\end{equation}

For $d=4$, corresponding to the observable universe, and for ordinary matter
one has

\begin{equation}
m^{(+)2}=\frac{2}{3}\Lambda ^{(+)}.  \label{81}
\end{equation}%
This mass expression must be associated with a systems traveling lower than
the light velocity $(c>v)$. In the case of particles traveling faster than
light velocity $(v>c)$, corresponding to tachyons, one obtains

\begin{equation}
m^{(-)2}=-\frac{2}{3}\Lambda ^{(-)}.  \label{82}
\end{equation}%
Note that since $\Lambda ^{(+)}>0$ and $\Lambda ^{(-)}<0$, both rest masses $%
m^{(+)2}$ and $m^{(-)2}$ are positives.

\bigskip \ 

\noindent \textbf{7. Linearized gravity in (4+6)-dimensions}

\smallskip \ 

The key idea in this section is to split the $(4+6)$-dimension as $%
(4+6)=(1+4)+(3+2)$. This means that the $(4+6)$-dimensional space is
splitted in two parts the de Sitter world of $(1+4)$-dimensions and anti de
Sitter world of $(3+2)$-dimensions. In this direction, let us write the line
elements in (3) and (7) in the form

\begin{equation}
d\mathcal{S}%
_{(+)}^{2}=-c^{2}dt_{0(+)}^{2}=-c^{2}dt_{(+)}^{2}+dx_{(+)}^{2}+dy_{(+)}^{2}+dz_{(+)}^{2}+dw_{(+)}^{2}
\label{83}
\end{equation}%
and

\begin{equation}
d\mathcal{S}%
_{(-)}^{2}=-c^{2}dt_{0(-)}^{2}=+c^{2}dt_{(-)}^{2}-dx_{(-)}^{2}-dy_{(-)}^{2}-dz_{(-)}^{2}+d\tau _{(-)}^{2},
\label{84}
\end{equation}%
respectively. One can drop the parenthesis notation in the coordinates $%
x_{(+)}^{\mu }$ and $x_{(-)}^{\mu }$ if one makes the convention that the
index $\mu $ in $x_{(+)}^{\mu }$ runs from $1$ to $4$, while the index $\mu $
in $x_{(-)}^{\mu }$ is changed to an index $a$ running from $5$ to $8$.
Thus, in tensorial notation one may write (83) and (84) as

\begin{equation}
d\mathcal{S}_{(+)}^{2}=\eta _{\mu \nu }^{(+)}dx^{\mu }dx^{\nu }+dw_{(+)}^{2}
\label{85}
\end{equation}%
and 
\begin{equation}
d\mathcal{S}_{(-)}^{2}=\eta _{ab}^{(-)}dx^{a}dx^{b}+d\tau _{(-)}^{2}.
\label{86}
\end{equation}%
Here, $\eta _{\mu \nu }^{(+)}$ and $\eta _{ab}^{(-)}$ denotes flat Minkowski
metrics with the signatures $(-1,+1,+1,+1)$ and $(+1,-1,-1,-1)$,
respectively. It seems evident that one can reach a unified framework by
adding (85) and (86) in the form

\begin{equation}
d\mathcal{S}^{2}=d\mathcal{S}_{(+)}^{2}+d\mathcal{S}_{(-)}^{2}=\eta _{\mu
\nu }^{(+)}dx^{\mu }dx^{\nu }+dw_{(+)}^{2}+\eta
_{ab}^{(-)}dx^{a}dx^{b}+d\tau _{(-)}^{2}.  \label{87}
\end{equation}

Let us assume that in a world of $(4+6)$-signature one has the two
constraints%
\begin{equation}
\eta _{\mu \nu }^{(+)}x^{\mu }x^{\nu }+w^{2}=\frac{3}{\Lambda _{(+)}}
\label{88}
\end{equation}%
and

\begin{equation}
\eta _{ab}^{(-)}dx^{a}dx^{b}+\tau ^{2}=\frac{3}{\Lambda _{(-)}},  \label{89}
\end{equation}%
where $\Lambda _{(+)}>0$ and $\Lambda _{(-)}<0$ again play the role of two
cosmological constants. Following similar procedure as in section 5,
considering the constraints (88) and (89) one can generalize the the line
element (87) in the form

\begin{equation}
d\mathcal{S}^{2}=g_{\; \; \; \; \mu \nu }^{(+)(0)}dx^{\mu }dx^{\nu }+g_{\;
\; \; \;ab}^{(-)(0)}dx^{a}dx^{b}.  \label{90}
\end{equation}%
where

\begin{equation}
g_{\; \; \; \; \mu \nu }^{(+)(0)}=\eta _{\mu \nu }^{(+)}+\frac{x_{\mu
}x_{\nu }}{\frac{3}{\Lambda _{(+)}}-x^{i}x^{j}\eta _{ij}^{(+)}}  \label{91}
\end{equation}%
and

\begin{equation}
g_{\; \; \; \;ab}^{(-)(0)}=\eta _{ab}^{(-)}+\frac{x_{a}x_{b}}{\frac{3}{%
\Lambda _{(-)}}-x^{i}x^{j}\eta _{ij}^{(-)}}.  \label{92}
\end{equation}

Using (91) and (92) one can define a background matrix $\gamma _{AB}$, with
indexes $A$ and $B$ running from $1$ to $8$ in the form

\begin{equation}
\gamma _{AB}^{(0)}=\left( 
\begin{array}{cc}
g_{\; \; \; \; \mu \nu }^{(+)(0)} & 0 \\ 
0 & g_{\; \; \; \;ab}^{(-)(0)}%
\end{array}%
\right) .  \label{93}
\end{equation}%
Thus, one can write the linearized metric associated with (93) as

\begin{equation}
\gamma _{AB}=\gamma _{AB}^{(0)}+h_{AB}.  \label{94}
\end{equation}%
Hence, following a analogous procedure as the presented in section 4, one
obtains the equation for $h_{AB}$ in $d=4+4=8$-dimensions,

\begin{equation}
\square ^{2}\bar{h}_{AB}-\frac{2}{21}\Lambda \bar{h}_{AB}=0.  \label{95}
\end{equation}%
Here, one has 
\begin{equation}
\square ^{2}=\gamma _{AB}^{(0)}\mathcal{D}^{A}\mathcal{D}^{B}.  \label{96}
\end{equation}%
One can split $\square ^{2}$ in the form%
\begin{equation}
\square ^{2}=\square ^{(+)^{2}}+\square ^{(-)^{2}},  \label{97}
\end{equation}%
where $\square ^{(+)^{2}}=g_{\; \; \; \; \mu \nu }^{(+)(0)}\mathcal{D}^{\mu }%
\mathcal{D}^{\nu }$ and $\square ^{(-)^{2}}=g_{\; \; \; \;ab}^{(0)(-)}%
\mathcal{D}^{a}\mathcal{D}^{b}$. One shall use now the separation of
variables method. For this purpose let us assume that the perturbation $\bar{%
h}_{AB}=$ $\bar{h}_{AB}(x^{\mu },x^{a})$ can be splitted in the form

\begin{equation}
\bar{h}_{AB}=\bar{h}_{\quad AC}^{(+)}(x^{\mu })\bar{h}_{\qquad
B}^{(-)C}(x^{a}).  \label{98}
\end{equation}%
Thus, one discovers that (95) becomes

\begin{equation}
\bar{h}_{\qquad B}^{(-)C}\square ^{(+)^{2}}\bar{h}_{\quad AC}^{(+)}+\bar{h}%
_{\quad AC}^{(+)}\square ^{(-)^{2}}\bar{h}_{\qquad B}^{(-)C}-\frac{2}{21}%
\Lambda \bar{h}_{\quad AC}^{(+)}\bar{h}_{\qquad B}^{(-)C}=0.  \label{99}
\end{equation}%
Multiplying the last equation by $\bar{h}^{(+)AE}\bar{h}_{\qquad D}^{(-)B}$
yields

\begin{equation}
\bar{h}^{(+)AE}\square ^{(+)^{2}}\bar{h}_{\quad AD}^{(+)}-\frac{2}{21}%
\Lambda \delta _{D}^{E}=-\bar{h}_{\qquad D}^{(-)B}\square ^{(-)^{2}}\bar{h}%
_{\qquad B}^{(-)E}.  \label{100}
\end{equation}%
Thus, one observes that while the left hand side of (100) depends only of $%
x^{\mu }$ and the right hand side depends only of $x^{a}$ one may introduce
a constant $\hat{\Lambda}$ such that

\begin{equation}
\bar{h}^{(+)AE}\square ^{(+)^{2}}\bar{h}_{\quad AD}^{(+)}-\frac{2}{21}%
\Lambda \delta _{D}^{E}=\hat{\Lambda} \delta _{D}^{E}  \label{101}
\end{equation}%
and

\begin{equation}
-\bar{h}_{\qquad D}^{(-)B}\square ^{(-)^{2}}\bar{h}_{\qquad B}^{(-)E}=\hat{%
\Lambda} \delta _{D}^{E}.  \label{102}
\end{equation}%
One may rewrite (101) and (102) in the form

\begin{equation}
\square ^{(+)^{2}}\bar{h}_{\quad AB}^{(+)}-\left( \frac{2}{21}\Lambda +\hat{%
\Lambda} \right) \bar{h}_{\quad AB}^{(+)}=0  \label{103}
\end{equation}%
and

\begin{equation}
\square ^{(-)^{2}}\bar{h}_{\quad AB}^{(-)}+\hat{\Lambda} \bar{h}_{\quad
AB}^{(-)}=0.  \label{104}
\end{equation}

According to the formalism presented in section 5, one can identify the
tachyonic mass in the anti-de Sitter-world by $m^{(-)2}=-\hat{\Lambda} $,
while the mass in the de\ Sitter-world by $m^{(+)2}=\frac{2}{21}\Lambda +%
\hat{\Lambda} $. Moreover, one can also introduce an effective $M^{2}=\frac{2%
}{21}\Lambda $ in the $(4+4)$-world. Note that the effective mass $M^{2}$
can be written as $M^{2}=m^{(+)2}+m^{(-)2}$. Thus, (103) and (104) can be
rewritten as

\begin{equation}
(\square ^{(+)^{2}}-m^{(+)2})\bar{h}_{\quad \mu \nu }^{(+)}=0  \label{105}
\end{equation}%
and

\begin{equation}
(\square ^{(-)^{2}}-m^{(-)2})\bar{h}_{\quad ab}^{(-)}=0.  \label{106}
\end{equation}%
Here, we fixed the gauges $\bar{h}_{\quad ab}^{(+)}=0$ and $\bar{h}_{\quad
\mu \nu }^{(-)}=0$. Thus, we have shown that from linearized gravity in $%
(4+4)$-dimensions one can derive linearized gravities in $(1+3)$-dimensions
and in $(3+1)$-dimensions. Moreover, the modes $\bar{h}_{\quad \mu \nu
}^{(+)}$ can be associated with a massive graviton in the de Sitter space,
while the modes $\bar{h}_{\quad ab}^{(-)}$ can be associated with the
tachyonic graviton in the anti-de Sitter space.

\ 

\ 

\ 

\noindent \textbf{8. Final remarks}

\smallskip \ 

In this work we have developed a higher dimensional formalism for linearized
gravity in the de Sitter or anti-de Sitter space-background which are
characterized by the cosmological constants $\Lambda ^{(+)}>0$ and $\Lambda
^{(-)}<0$, respectively. Our starting point are the higher dimensional
Einstein gravitational field equations and the perturbed metric $g_{\mu \nu
}^{(\pm )}=g_{\; \; \; \; \mu \nu }^{(\pm )(0)}+h_{\mu \nu }^{(\pm )}$,
where $g_{\; \; \; \; \mu \nu }^{(\pm )(0)}$ is a background metric
associated with the cosmological constants $\Lambda ^{(+)}$, $\Lambda ^{(-)}$
and the Minkowski flat metric $\eta _{\; \; \; \; \mu \nu }^{(\pm )}$. After
straightforward computations and after imposing a gauge conditions for $%
h_{\mu \nu }^{(\pm )} $ we obtain the two equations (71) and (72). We proved
that these two equations admit an interpretation of massive graviton with
mass given by (79) and (80). According to the formalism discussed in section
2, the massive graviton with mass $m^{(+)2}$ can be associated with ordinary
graviton which \textquotedblleft lives\textquotedblright \ in the de Sitter
space, while the massive graviton with mass $m^{(-)2}$ is a tachyonic
graviton which \textquotedblleft lives\textquotedblright \ in the anti-de
Sitter space. We should mention that these results agree up to sign from
those described by Novello and Neves [17]. The origin of this difference in
the signs is that although they consider a version of linearized gravity
their approach refers only to four dimensions and rely in a field strength $%
F_{\mu \nu \alpha \beta }$ which is not used in our case. Here, we get a
four dimensional graviton mass $m_{g}^{2}=\frac{2}{3}\Lambda $ for de\
Sitter space and using the Planck 2015 data\ [18] we can set $m_{g}\sim
3.0\times 10^{-69}kg$, while the current upper bound obtained by the
detection of gravitational waves is $m_{g}\leq 1.3\times 10^{-58}kg$ [15].

Furthermore, in the previous section, we discuss the case of the $(4+6)$
signature where we identify $m^{(+)2}$ and $m^{(-)2}$ as a contribution to
an effective mass $M^{2}$ in the unified framework of $(4+4)$-dimensions. It
would be interesting for a future work to have a better understanding of the
meaning of the mass $M^{2}$. Also, it may be interesting to extent this work
to a higher dimensional cosmological model with a massive graviton.

On the other hand, it is worth mentioning that our proposed formalism in $%
(4+4)$-dimensions may be related to the so called double field theory [19].
This is a theory formulated with $x^{A}=(x^{\mu },x^{a})$ coordinates
corresponding to the double space $R^{4}\times \ T^{4}$, with $A=1,2,...,8$
and $D=8=4+4$. In this case the constant metric is given by

\begin{equation}
ds^{2}=\eta _{AB}dx^{A}dx^{B}.  \label{107}
\end{equation}%
Moreover, the relevant group in this case is $O(4,4)$ which is associated
with the manifold $M^{8}$. It turns out that $M^{8}$ can be compactified in
such a way that becomes the product $R^{4}\times \ T^{4}$ of flat space and
a torus. In turn the group $O(8,8)$ is broken into a group containing $%
O(4,4)\times O(4,4;Z)$. A detail formulation of this possible relation will
be present elsewhere.

Finally, it is inevitable to mention that perhaps the formalism developed in
this work may be eventually useful for improvements of the direct detection
of gravitational waves. This is because recent observations [20] established
that the cosmological value has to be small and positive and that the
observable universe resembles to a de Sitter universe rather than an anti de
Sitter universe. Also, it will be interesting to explore a link between this
work and the electromagnetic counterpart of the gravitational waves [21].

\ 

\begin{center}
\textbf{Acknowledgments}

\smallskip
\end{center}

\ We would like to thank to E. A. Le\'{o}n for helpful comments. We would
also like to thank the referee for valuable comments. This work was
partially supported by PROFAPI 2013.\newpage 

\noindent \textbf{Appendix A. Negative mass squared term - tachyon
association}

\smallskip

\smallskip

This appendix is dedicated to clarify why the expression (80) refers to a
tachyon system. In some sense the below presentation is the reverse argument
as the one presented in section 2.

Consider the Klein-Gordon equation

\begin{equation}
({\square }^{2}+m_{0}^{2})\varphi =0.  \label{A1}
\end{equation}%
If one considers a plain wave solution for (A1), the solution can be written
as 
\begin{equation}
\varphi =Ae^{p^{\mu }x_{\mu }},  \label{A2}
\end{equation}%
where $A$ is a constant. Therefore, using (A2) one can verify that (A1) is
reduced to 
\begin{equation}
(p^{2}+m_{0}^{2})\varphi =0,  \label{A3}
\end{equation}%
which implies that 
\begin{equation}
p^{2}+m_{0}^{2}=0.  \label{A4}
\end{equation}%
Since $p^{\mu }=m_{0}u^{\mu }$ one discovers that (A4) leads to a relation
of the form 
\begin{equation}
dt=\frac{d\tau }{\sqrt{1-v^{2}/c^{2}}}.  \label{A5}
\end{equation}%
Which implies that $v<c$ and therefore the system moves with velocities less
than the light velocity. Similarly if instead of (A1) one considers the
expression 
\begin{equation}
({\square }^{2}-m_{0}^{2})\varphi =0.  \label{A6}
\end{equation}%
A plain wave solution would imply the classical expression 
\begin{equation}
{p}^{2}-m_{0}^{2}=0,  \label{A7}
\end{equation}%
which again considering the relation $p^{\mu }=m_{0}u^{\mu }$ one finds that
instead of the relation (A5) one has 
\begin{equation}
dt=\frac{d\tau }{\sqrt{v^{2}/c^{2}-1}}.  \label{A8}
\end{equation}%
This implies that $v>c$ and therefore describes a tachyon system. If instead
of the field $\varphi $ one considers the $h_{\mu \nu }$ and assume a plain
wave solution of the form $h_{\mu \nu }=A_{\mu \nu }e^{p^{\mu }x_{\mu }}$,
one may be able to obtain the corresponding expression (A5) and (A8) for
linearized gravity.

\end{document}